\documentclass{PoS}

\title{Field theory for biophysical neural networks}

\ShortTitle{Field theory for neural networks}

\author{\speaker{Siwei Qiu}\\
        Laboratory of Biological Modeling, NIDDK, NIH, Bethesda, Maryland, U.S.A\\
        E-mail: \email{siwei.qiu@gmail.com}}

\author{Carson Chow\\
        Laboratory of Biological Modeling, NIDDK, NIH, Bethesda, Maryland, U.S.A\\
        E-mail: \email{carsonc@mail.nih.gov}}

\abstract{The human brain is a complex system composed of a network of
hundreds of billions of discrete neurons that are coupled through time dependent synapses.  Simulating the entire brain is a daunting challenge.  Here, we show how
ideas from quantum field theory can be used to construct an effective reduced theory, which may be analyzed with lattice computations.  We give some examples of how the formalism can be applied to biophysically plausible neural network models.}

\FullConference{The 32nd International Symposium on Lattice Field Theory\\
                 23-28 June, 2014\\
                 Columbia University New York, NY}

\begin{document}

\section{Introduction\label{sec:intro}}
The human brain is the most complex system in the known universe and we only have a primitive understanding of how it functions.  It is a dense collection of cells and tissue.  The functional elements are cells called neurons of which there are approximately 
$10^{11}$.  Each neuron is itself a complex electrochemical cell with many types of voltage-dependent ion channels on its membrane, which is an insulating lipid bilayer.  Over a half century ago, Hodgkin and Huxley~\cite{HodgkinHuxley,Morris1981,Connor1971} showed that the neuron membrane dynamics could be modeled as a nonlinear electrical RC circuit, where the membrane acts like a capacitor and the ion channels act like nonlinear resistors (i.e. conductances) which can be written in the form
\begin{equation}
C\frac{dV}{dt} = -\sum_i g_i (V-V_i^0)
\end{equation}
where $V$ is the membrane voltage, $C$ is the capacitance of the membrane, $g_i$ is a nonlinear and possibly time-dependent
 voltage-dependent conductance, and $V_i^0$ is a baseline reversal potential for the channel.  
For low voltages, the circuit is at rest but if a higher voltage (or current) is applied to the membrane, 
then an action potential or spike is triggered where the voltage rises abruptly and then returns to rest. 
 Neurons are connected to each other through synapses, which are molecular machines that exchange chemical neurotransmitters.
A spike in an upstream neuron induces neurotransmitter to be released at the synapse, which is received by a receptor in the downstream neuron.  The receptor induces ion channels to open in the downstream neuron that can increase or decrease the voltage.  An increase (called excitation) makes a spike more likely while a decrease (called inhibition) makes a downstream spike less likely.  
The neuron consists of a cell body, external input structures called dendrites and an extended output wire called an axon.  The synapses are located at the junction between axons and dendrites.  A detailed biophysical model of a neuron itself could consist of thousands of differential equations.  Thus it is easy to see that a simulation of the entire brain is a great challenge.

In order to make theoretical progress, some amount of reduction is required.  When the neuron is near the threshold to spiking, the neuron dynamics can be well modeled by only keeping track of the phase of the voltage~\cite{Ermentrout95typei}.  This is called the theta neuron model. We consider a network of theta neurons described by\begin{eqnarray}
\dot{\theta}_i&=&1-\cos\theta_i+(I_i+u_i-(D/2)\sin\theta_i+\sqrt{D}dW/dt)(1+\cos\theta_i)\label{eq:micro1}\\
\dot{u}_i&=&-\beta u_i+2\beta\frac{1}{N}\sum_{j=1}^Nw_{ij}\eta_j(\pi,t)\label{drive}\\
\eta_i(\theta,t)&=&\delta(\theta-\theta_i(t))\label{eq:micro3}
\end{eqnarray} 
where $\theta_i$ is the phase of a neuron at spatial location $i$, $u_i$ is the net "synaptic drive" to neuron $i$, $w_{ij}$ is the coupling weight between sites $i$ and $j$, $N$ is the total number of neurons,  $\eta_i(\theta,t)$ is the neuron phase density at $i$,  and $dW$ is white noise forcing with variance $D$.  The neurons are considered to spike when their phase is $\pi$, which induces a "kick" to the synaptic drive that is low-pass filtered with time constant $\beta$.  There is a factor of two in (\ref{drive}) because the phase velocity at $\pi$ is 2.  We are interested in the various possible dynamical states for this microscopic system.  In particular, we would like to understand the limit of large but finite $N$.  We show how the formalism from quantum field theory can be used to represent this system as a path integral from which perturbation theory can be applied.  In particular, we can derive an effective action for the mean and higher moments (defined in probability theory) of the system.

\section{General action and mean field equation}
It has been shown previously~\cite{FSE2013}, that the probability density functional for the phase density and synaptic drive  can be expressed as a path integral
\begin{equation}
P[\vec\phi (\theta,t),\vec u(t)]=\int [d\vec\phi][d \vec u]  e^{-S}
\end{equation}
with action  $S=S_{\phi}+S_u$, where
\begin{eqnarray}
S_{\phi}&=&\int d\theta dt\sum_i M_i\tilde{\phi}_i(\theta,t)[\partial_t\phi_i(\theta,t)+\partial_{\theta}F(\theta,u_i)\phi_i(\theta,t)-\frac{1}{2}\partial_{\theta}^2 D(1+\cos\theta)^2\phi_i(\theta,t)]\\
S_u&=&\int dt\sum_i\tilde{u}_i\bigg(\frac{d}{dt}u_i(t)+\beta u_i-2\beta\frac{1}{N}\sum_j w_{ij}(1+\tilde{\phi}_j(\pi,t))\phi_j(\pi,t)\bigg)
\end{eqnarray}
We have applied the Doi-Peliti-Janssen transformation, $\phi=\eta e^{-\tilde{\eta}}$, $\tilde{\phi}=e^{\tilde{\eta}}-1$
and  $\tilde{u}$ and $\tilde{\eta}$ are auxiliary fields.
The action was derived by noting that formally the probability density functional can be expressed as a Dirac delta functional of the phase density and synaptic drive constrained to 
Eq.~\ref{drive} and the continuity equation for the density, Eq.~\ref{eq:micro3}.  The path integral then arises by making use of
the Fourier decomposition of the Dirac delta functional with the auxiliary fields as the Fourier transform variables.

Mean field theory (equations of motion) obtained by minimizing the action is 
\begin{eqnarray}
&&\frac{d}{dt}u_i(t)+\beta u_i(t)-2\beta\frac{1}{N}\sum_j w_{ij}\phi_j(\pi,t)=0\label{eq:mean1}\\
&&\partial_t\phi_i(\theta,t)+\partial_{\theta}[1-\cos\theta+(I_i(t)+\gamma u_i(t)-D/2\sin\theta)(1+\cos\theta)]\phi_i(\theta,t)\nonumber\\
&&-\frac{1}{2}\partial^2_{\theta}D(1+\cos\theta)^2\phi_i(\theta,t)=0\label{eq:mean2}
\end{eqnarray}

Mean field theory describes the contribution from the first moment (defined in probability theory) and is valid in the
limit of $N\rightarrow\infty$.  
Similar to interacting particles in a gas, pairwise correlations between neurons can produce statistical fluctuations 
that are proportional to $1/N$.
Mean field theory represents the ideal gas, while higher moment models describe the 
interacting gas, where finite $N$ effects can play a role.  Real neural networks have large but finite $N$ and we are interested when mean field theory is applicable and when it breaks down due to finite $N$ effects.

\section{Uniform coupling~\label{sec:uniform}}
The simplest coupled network is one where the coupling is uniform,  $w_{ij}=w_0$.  
Although this is not a realistic model, it is useful for developing the theoretical methodology.
In this example, we also assume that the system has no noise, which means setting $D=0$. 
Since each neuron receives the same input, we do not need a site index. 
Consequently, the mean field equations are reduced to:
\begin{eqnarray}
&&(\frac{d}{dt}+\beta)u(t)-2\beta w_0 \phi(\pi,t)=0\label{eq:meanu}\\
&&(\partial_t+\partial_{\theta}F(\theta,u(t)))\phi(\theta,t)=0\label{eq:meanphi}
\end{eqnarray}
where $F(\theta,u(t))=1-\cos\theta+(1+\cos\theta)(I+u(t))$. Also, for simplicity, we set $I$ to be a constant. 

We want to understand the dynamics of the mean field equations.  In particular, we want to study the dynamical attractors of the system.
The 
stationary solution is 
\begin{eqnarray}
u_{0}&=&\frac{w_0^2}{2\pi^2}\bigg(1\pm\sqrt{1+\frac{4I\pi^2}{w_0^2}}\bigg)\label{u0}\\
\phi_{0}(\theta)&=&\frac{\sqrt{I+u_{0}}}{\pi F(\theta,u_{0})} \label{phi0}
\end{eqnarray} 
where the branch of $u_0$ to be selected is the one that matches the same sign as $w_0$.  The solution $\phi_{0}(\theta)$ is only valid when $I+u_{0}>0$. 
When $I+u_{0}<0$, $\phi_{0}=\delta(\theta-\theta_0)$, where $\theta_0$ is given by $F(\theta_0,u_{0})=0$. 
Here, we will only consider the case of $I+u_{0}>0$. 

Consider the perturbation
\begin{eqnarray}
u(t)&=&u_{0}+\epsilon v(t)\label{eq:pertu}\\
\phi(\theta,t)&=&\phi_{0}(\theta)+\epsilon\eta(\theta,t)\label{eq:pertphi}
\end{eqnarray}
Set $v(t)=e^{\lambda t}C_v$ and substitute into equations~ \ref{eq:pertu}  and \ref{eq:meanu} to obtain
\begin{eqnarray}
\eta(\pi,t)=e^{\lambda t}C_v\frac{\lambda+\beta}{2\beta w_0}\label{eq:eta2}
\end{eqnarray}
Substituting $\eta(\theta,t)=e^{\lambda t}b(\theta)$ into equations~\ref{eq:pertphi} and~\ref{eq:meanphi}, 
gives the ordinary differential equation (ODE):
\begin{eqnarray}
\frac{d}{d\theta}b(\theta)+p(\theta)b(\theta)=q(\theta)
\end{eqnarray}
where
\begin{eqnarray}
p(\theta)&=&\frac{\lambda+\sin\theta(1-I-u_0)}{F(\theta,u_0)}\\
q(\theta)&=&\frac{2C_v\sqrt{I+u_0}\sin\theta}{\pi F(\theta,u_0)}
\end{eqnarray}
We solve this ODE to get $b(\theta)$, from which we obtain $b(\pi)$ to substitute into \ref{eq:eta2}.  This then results in 
an equation for $\lambda$:
\begin{eqnarray}
\lambda^3+\beta\lambda^2+\gamma\lambda+\beta\gamma-\alpha=0
\end{eqnarray}

Here, we have $\gamma=4(I+a_{10})$ and $\alpha=2\beta w_0\sqrt{I+u_0}/\pi$. 
The solutions of this cubic equation of $\lambda$ determine stability of the stationary solution in (\ref{phi0}).
When $\lambda$ has positive real part, 
the system is unstable, and when it has negative real part, the system is stable. 
For this simple model, when $w_0$ is positive, the stationary solution is stable, 
and when $w_0$ is negative, it is unstable. Furthermore, when $w_0$ is negative, the 
neural oscillators will synchronize. In this case,
the population function $\phi(\theta,t)$ will coalesce into a rotating delta function pulse.
This is difficult to simulate numerically since the angle derivatives will become unbounded.
The way to avoid this 
difficulty is to introduce noise, which will regularize the equation, similar to the technique of smearing in lattice $QCD$.

\section{Mexican hat coupling case}
\begin{figure}[h]
    \centerline{
 \includegraphics[width=0.33\linewidth]{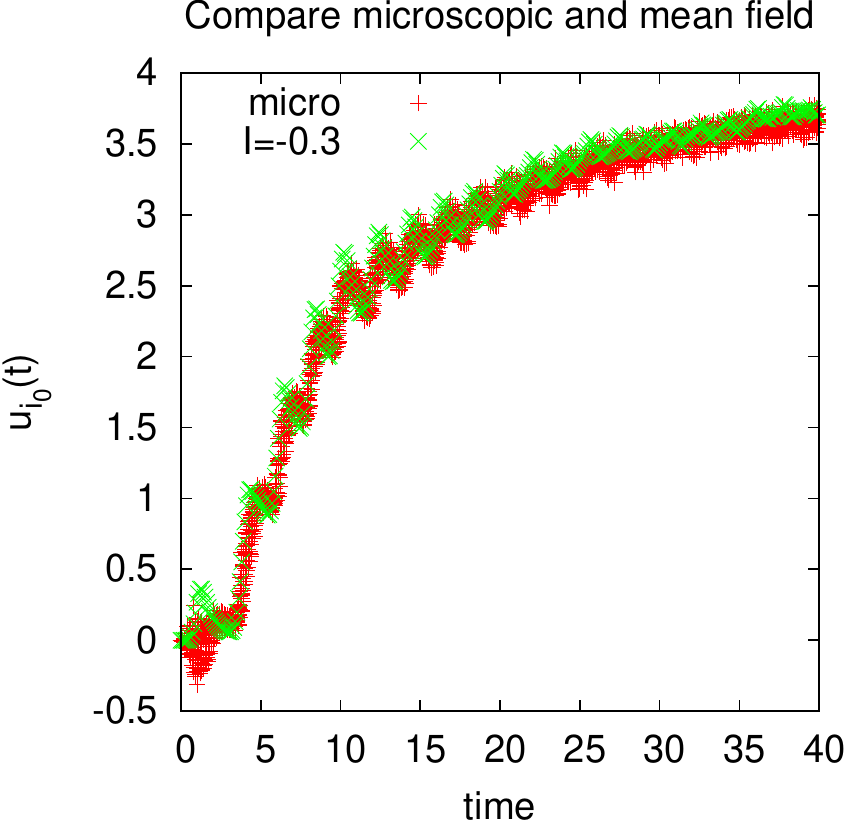}%
 \includegraphics[width=0.36\linewidth]{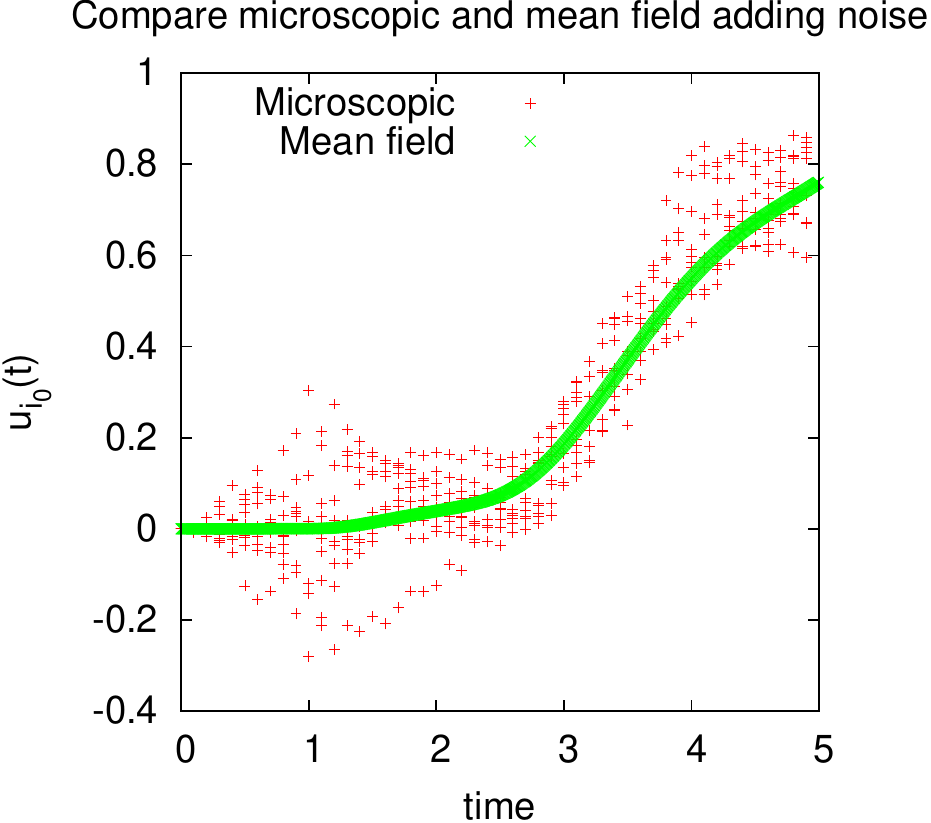}\\%
}
    \caption{Comparison $u_{i_0}(t)$ of mean field equation and microscopic equation . We are setting
$\beta$ as 0.2 here. Left: without noise. Right: with noise}
    \label{fig:compare}
\end{figure}

The next level of complexity is include spatial dependence. 
We investigated a "Mexican hat" coupling weight, where the coupling coefficient is defined as:
\begin{eqnarray}
w_{ij}=A\exp[-a|i-j|dz]-\exp[-|i-j|dz]
\end{eqnarray}
Here, $A$ and $a$ are constant, which are both set to be $2.6$ in the simulation. $i$ and $j$ are the site of two neurons
within the network. $dz$ is the lattice constant.  
We discretize the differential operator in Eq.~\ref{eq:mean1} and Eq.~\ref{eq:mean2} for numerical simulations. 
We tried different numerical schemes, including 
forward Euler, backward Euler, and a symplectic scheme, which all performed well.
From the result of these simulations, we find that when $\beta$ is small, the mean field system matches the microscopic system
but for large $\beta$, the mean field system deviates from the microscopic system. 
Fig.~\ref{fig:compare} shows the 
result of $u_{i_0}(t)$ in mean field(Eq.~\ref{eq:mean1} and Eq.~\ref{eq:mean2}) and 
microscopic equations (Eq.~\ref{eq:micro1} to Eq.~\ref{eq:micro3}) for small $\beta$. 
Here, $i_0$ is the site in the middle of the spatial domain.

\begin{figure}
\centerline{
\includegraphics[width=0.33\linewidth]{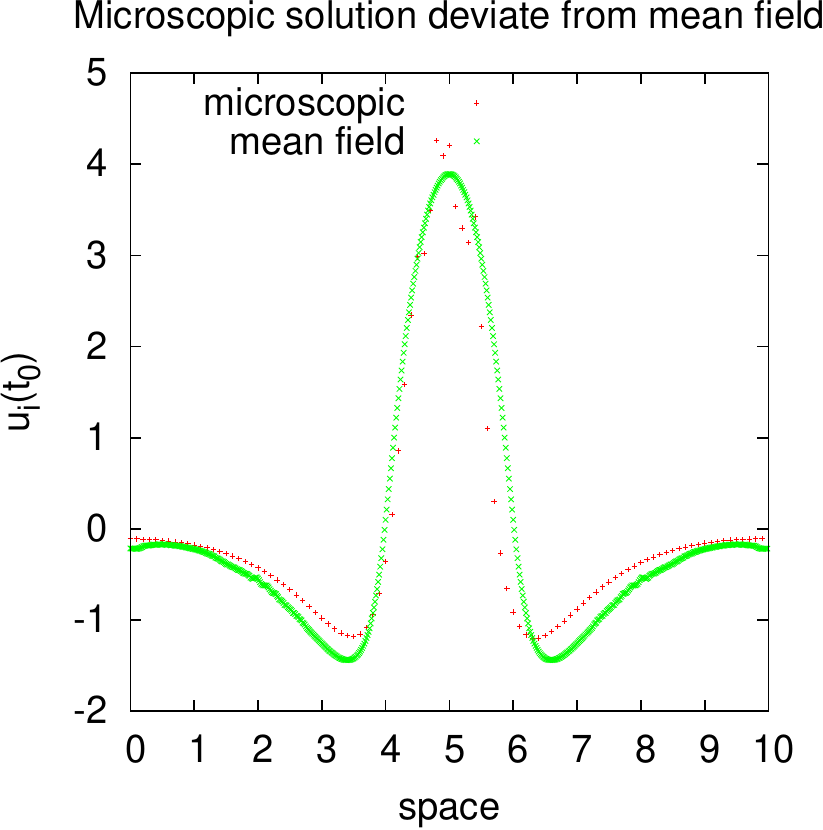}
}
\caption{Comparison of $u_i(t_0)$ of mean field equation and microscopic equation at fixed time $t_0$. $\beta=3$.}
 \label{fig:dive}
\end{figure}

In Fig.~\ref{fig:dive}, we show the comparison for large $\beta$, where the mean field and microscopic results deviate.  
This deviation is a sign of a finite $N$ effect, which is the contribution of higher moments. One important question is how 
to characterize these contributions systematically. We will show our methodology in the next section.

\section{Preliminary analysis of finite $N$ effect}
The loop expansion~\cite{FSE2013} and 
2PI (activity) equations~\cite{2pi1974, activity2013} are possible methods for analyzing finite $N$ effects. Here, we give some results for the activity equations. The activity  equations are obtained by 
performing a Legendre transform from our original fields ($u_i(t)$ and $\phi_i(\theta,t)$) 
to moment variables. Here, we will focus on the dynamics that includes the contribution from two-point correlation functions. 

Following  the notation of~\cite{activity2013}, we denote $\xi_1(y)=u(z,t)$, $\xi_2(y)=\tilde{u}(z,t)$, $\xi_3(x)=\phi(x)$ 
and $\xi_4(x)=\tilde{\phi}(x)$. Here, $y=(z,t)$ and $x=(z,t,\theta)$. We use variable $z$ instead of site index. 
We also denote mean value $a_i=\langle \xi_i\rangle$ and correlation matrix $C_{ij}=N[\langle \xi_i\xi_j\rangle-a_ia_j]$. 
Here, $i$ and $j$ are not site index, but set $\{1,2,3,4\}$. The activity equations for our model are then written as:
\begin{eqnarray}
&&\frac{d}{dt}a_1(z,t)+\beta a_1(z,t)-2\beta\int dz^{\prime}w(z-z^{\prime})a_3(\pi,z^{\prime},t)=0\\
&&\frac{d}{dt}a_3(\theta,z,t)+\partial_{\theta}F(\theta,a_1(z,t))a_3(\theta, z, t)-\frac{1}{2}\partial_{\theta}^2D(1+\cos\theta)^2a_3(\theta,z,t)\nonumber\\
&&+\frac{\gamma}{2N}\partial_{\theta}(1+\cos\theta)(C_{31}(\theta,z,t;z,t)+C_{13}(z,t;z,t,\theta))=0\\
&&(\frac{d}{dt}+\beta)C_{11}(z,t;z_0,t_0)-2\beta\int dz^{\prime}w(z-z^{\prime})C_{31}(\pi,z^{\prime},t;z_0,t_0)=0\\
&&(\frac{d}{dt}+\beta)C_{13}(z,t;\theta_0,z_0,t_0)-2\beta\int dz^{\prime}w(z-z^{\prime})C_{33}(\pi,z^{\prime},t;\theta_0,z_0,t_0)-2\beta w(z-z_0)a_3(\pi,z_0,t_0)=0\nonumber\\
&&\\
&&(\partial_{\theta}\gamma(1+\cos(\theta)))a_3(x)C_{11}(z,t;z_0,t_0)+[\partial_t+\partial_{\theta}F(\theta,a(z,t))\nonumber\\
&&-\frac{1}{2}\partial_{\theta}^2D(1+\cos\theta)^2]C_{31}(\theta,z,t;z_0,t_0)-2\beta w(z-z_0)a_3(\pi,z_0,t_0)\delta(\pi-\theta)=0\\
&&(\partial_{\theta}\gamma(1+\cos(\theta)))a_3(x)C_{13}(z,t;\theta_0,z_0,t_0)\nonumber\\
&&+[\partial_t+\partial_{\theta}F(\theta,a(z,t))-\frac{1}{2}\partial_{\theta}^2D(1+\cos\theta)^2]C_{33}(\theta,z,t;\theta_0,z_0,t_0)=0
\end{eqnarray}

Here, $F(\theta,a_1(z,t))=1-\cos\theta+(I(z,t)+\gamma a_1(z,t)-D/2\sin\theta)(1+\cos\theta)$.
These activity equations contain information about correlations, so it is a first order correction of the mean field equations.
Various analyses of this system is ongoing. First, we are analyzing the stability of the steady state solutions as
in section~\ref{sec:uniform}. 
Second, we are searching for efficient numerical algorithms for simulation of activity equations. Results will 
be reported in future work. 

\section{Discussion and future direction}
In conclusion, we are developing a systematic way to analyze realistic neural networks analytically and numerically.
In order to get an efficient algorithm for numerical analysis, we first need to analyze the stability of the solutions in different
parameter regimes.  For example, stability of a state is a necessary condition for the neural network to hold a memory.
From the shape of $u(t)$ in Fig.~\ref{fig:dive}, we see that there is a bounded region where $u(t)$ is positive. 
So the uniform coupling stability analysis allows us to see that the region where $u(t)$ is positive is stable. 
However, the region where $u(t)$ is negative is only stable because there exist the boundary conditions, 
which will constrain the eigenvalues of the differential equation. Simultaneously, we know that
for the real neural network, $u(t)$ is not always stable, as can be seen in the simulation of microscopic system. 
The boundary where $u(t)=0$ 
is not stable, allowing $u(t)$ to wander. This is also what we see in Fig.~\ref{fig:dive}.
Our simulations show that  the stationary mean field solution of $u(t)$ is stable. So in the future, we should focus on the contribution 
from the correlation functions.
Hence, it is important to analyze the activity equations to see whether the dynamics of the correlation functions can describe the
system better. All in all, our next step is develop a lattice model which is easy to perform numerical simulations for the
purpose of probing the contribution of two point correlation function. 

This work was supported by the Intramural Research Program of the NIH, NIDDK.

\bibliographystyle{ieeetr}
\bibliography{NeuroProceeding}

\begin{thebibliography}{1}

\bibitem{HodgkinHuxley}
A.~L. Hodgkin and A.~F. Huxley, ``A quantitative description of membrane
  current and its application to conduction and excitation in nerve.,'' {\em
  The Journal of Physiology}, vol.~117, pp.~500--544, 1952.

\bibitem{Morris1981}
C.~Morris and H.~Lecar, ``Voltage oscillations in the barnacle giant muscle
  fiber,'' {\em Biophysical Journal}, vol.~35(1), pp.~193--213, 1981.

\bibitem{Connor1971}
C.~J. A and S.~C. F., ``Prediction of repetitive firing behaviour from voltage
  clamp data on an isolated neurone soma.,'' {\em The Journal of Physiology},
  vol.~213(1), pp.~31--53, 1971.

\bibitem{Ermentrout95typei}
B.~Ermentrout, ``Type i membranes, phase resetting curves, and synchrony,''
  {\em Neural Comput}, vol.~8, pp.~979--1001, 1995.

\bibitem{FSE2013}
M.~A. Buice and C.~C. Chow, ``Dynamic finite size effects in spiking neural
  networks,'' {\em PLoS Comput Biol}, vol.~9, p.~e1002872, 01 2013.

\bibitem{2pi1974}
J.~M. Cornwall, R.~Jackiw, and E.~Tomboulis, ``Effective action for composite
  operators,'' {\em Phys. Rev. D}, vol.~10, pp.~2428--2445, Oct 1974.

\bibitem{activity2013}
M.~A. Buice and C.~C. Chow, ``Generalized activity equations for spiking neural
  network dynamics,'' {\em Frontiers in Computational Neuroscience}, vol.~7,
  no.~162, 2013.

\end{thebibliography}

\end{document}